\documentclass[prl,aps,twocolumn,preprintnumbers,showpacs,showkeys,superscriptaddress]{revtex4}

\usepackage{epsfig,amssymb,amsfonts,verbatim}

\def\bfl{\begin{flushleft}}
\def\efl{\end{flushleft}}
\def\bfr{\begin{flushright}}
\def\efr{\end{flushright}}
\def\bc{\begin{center}}
\def\ec{\end{center}}
\def\be{\begin{equation}}
\def\ee{\end{equation}}
\def\ba{\begin{eqnarray}}
\def\ea{\end{eqnarray}}
\def\baa#1{\begin{array}{#1}}
\def\eaa{\end{array}}
\def\bw{\begin{widetext}}
\def\ew{\end{widetext}}
\def\nn{\nonumber }
\def\lb#1{\label{#1}}

\def\text#1{\mbox{#1}}

\begin{document}


\title{Diagnosis of transport properties in Ferromagnets}

\author{Andrew Das Arulsamy}

\address{Condensed Matter Group, Division of Exotic Matter, No. 22, Jalan Melur 14,
Taman Melur, 68000 Ampang, Selangor DE, Malaysia}


\date{\today}


\begin{abstract}

The ionization energy based resistivity model with further
confinements from spin-disorder scattering and polaronic effect is
derived so as to capture the mechanism of both spin independent
and spin-assisted charge transport in ferromagnets. The computed
$T_{crossover}$ below $T_C$ and carrier density in
Ga$_{1-x}$Mn$_{x}$As ($x$ = 6-7\%) are 8-12 K and 10$^{19}$
cm$^{-3}$, identical with the experimental values of 10-12 K and
10$^{18}$-10$^{20}$ cm$^{-3}$ respectively. The calculated charge
densities for Mn$_{0.02}$Ge$_{0.98}$ and La$_{1-x}$Ca$_{x}$MnO$_3$
($x$ = 10-20\%) are 10$^{19}$ cm$^{-3}$ and 10$^{17}$ cm$^{-3}$
respectively.
\end{abstract}

 \maketitle



\subsection{\textbf{1. Introduction}}\lb{s-in}

Diluted magnetic semiconductors (DMS) have the tremendous
potential for the development of spintronics and subsequently will
lay the foundation to realize quantum computing. In order to
exploit the spin assisted charge transport, one needs to
understand the transport mechanism such as the variation of
resistivity with temperature and doping in both paramagnetic and
ferromagnetic phases. A wide variety of the magneto-electronic
properties based on doping and Mn's valence state in manganites
were reported to understand the transport
mechanism(s)~\cite{horyn,chan}. Among them, the influence of grain
boundary as a barrier~\cite{ju}, as a region of depleted
$T_C$~\cite{hernandez} and polaronic effect~\cite{yang} on
electrical properties were reported. Direct proportionality of
{\bf H} with $T_C$~\cite{abramovich} and resistivity with defects
or substrate-film lattice
incompatibility~\cite{heffner,prokhorov,sun,mei,coldea} are also
regarded as equally important to determine the electrical
properties of manganites. Furthermore, metallic conduction below
$T_C$ has been studied using double exchange mechanism (DEM)
between $s$ and $d$ orbitals~\cite{demin} and the displacement of
hysteresis loop in field-cooled sample with an additional scenario
of non-linear spin and charge fluctuations due to
magnon~\cite{solontsov}. Explanations in term of hopping electrons
and DEM~\cite{yu}, and the influence of microstructural transition
arises from ionic radius or valence state of Nd in
Nd$_x$Sm$_{1-x}$Ca$_{0.8}$MnO$_3$~\cite{filippov} were also
reported extensively. The effect of hydrostatic (external)
pressure ($P$ = 0 $\to$ 15 kbar) and chemical doping (internal
$P$) on metal-insulator transition of Pr-Ca,La-Sr-MnO$_3$ have
been reported as well.~\cite{medvedeva}.

Interestingly, Van Esch {\it et al}.~\cite{esch1} have proposed
multiple exchange interactions, which are ferromagnetic (FM)
hole-hole and antiferromagnetic (AFM) Mn-hole interactions for
DMS. These two effects, after neglecting the direct exchange
between Mn-Mn (due to very diluted nature of DMS) are seem to be
sufficient enough to describe the temperature dependent
magnetization curves ($M(T)$) accurately. However, even after
inclusion of FM and AFM effects including the spin disorder
scattering, the transport property in the FM phase is still not
well understood. Unfortunately, this is also true for the case of
metallic property below $T_C$ in the well known and extensively
studied FM manganites as pointed out by Mahendiran {\it et
al}.~\cite{mahendiran2}. The resistivity ($\rho(T)$) above $T_C$
for manganites is found to be in an activated form described by
the equation~\cite{mahendiran2},

\begin {eqnarray}
\rho(T>T_C) = \rho_0\exp\bigg(\frac{E_a}{k_BT}\bigg).\label{eq:1}
\end {eqnarray}

$E_a$ is the activation energy, $\rho_0$ and $k_B$ denote the
residual resistivity at $T$ $\gg$ $E_a$ and Boltzmann constant
respectively. In the FM phase, the influence of $M(T)/M_0$ is more
pronounced than the electron-phonon ({\it e-ph}) contribution
where the latter requires an overwhelmingly large coupling
constant~\cite{mahendiran2}. Note that $M_0$ is the magnitude of
magnetization at 0 K. Therefore, Mahendiran {\it et al}. have
suggested that conventional mechanism namely, {\it e-ph}
scattering has to be put aside so as to explain the $\rho(T)$ for
manganites below $T_C$. On the contrary, $\rho(T)$ with {\it e-ph}
involvement for DMS in the paramagnetic phase is given
by~\cite{esch1}

\begin {eqnarray}
\rho(T>T_C) = \frac{C_1 +
C_2\big[\exp\big(\Theta_D/T\big)-1\big]^{-1}}{k_BT\ln\big[1 +
\exp\big((E_m - E_f)/k_BT\big)\big]}.\label{eq:2}
\end {eqnarray}

The term, $C_2/\big[\exp(\Theta_D/T)-1\big]$ takes care of the
{\it e-ph} contribution. $\Theta_D$, $E_f$, $E_m$, $C_1$ and $C_2$
represent the Debye temperature, Fermi level, mobility edge and
numerical constants respectively. The $\rho(T)$ in the FM phase
based on the spin disorder scattering as derived by
Tinbergen-Dekker is given by~\cite{tinbergen3}

\begin {eqnarray}
&&\rho_{SD}(T<T_C) = \frac{(m^*_{e,h})^{5/2}N(2E_F)^{1/2}}{\pi
(n,p) e^2\hbar^4}J_{ex}^2 \nn \\&& \times \bigg[S(S+1) -
S^2\bigg(\frac{M_{TD}(T)}{M_0}\bigg)^2 -
S\bigg(\frac{M_{TD}(T)}{M_0}\bigg) \nn \\&& \times \tanh
\bigg(\frac{3T_CM_{TD}(T)}{2TS(S+1)M_0}\bigg)\bigg]. \label{eq:3}
\end {eqnarray}

$N$ is the concentration of nearest neighbor ions (Mn's
concentration) while ($n,p$) is the concentration of charge
carriers (electrons or holes respectively). $m_{e,h}^*$ denotes
effective mass of electrons or holes, $\hbar$ = $h/2\pi$, $h$ =
Planck constant. $e$ is the charge of an electron, $E_F$ and
$J_{ex}$ are the Fermi and FM exchange interaction energies
respectively while $S$ is the spin quantum number.
Equation~(\ref{eq:3}) becomes equivalent to Kasuya~\cite{kasuya4}
if one replaces the term,
$\tanh\big[3T_CM_{TD}(T)/2TS(S+1)M_0\big]$ with 1. Again, an
accurate equation for the $\rho(T)$ below $T_C$ is still lacking
since spin disorder scattering alone is insufficient as shown by
Tinbergen and Dekker~\cite{tinbergen3} as well as reviewed by
Ohno~\cite{ohno5}.

As a consequence, it is desirable to derive a formula that could
describe the transport mechanism of ferromagnets for the whole
temperature range i.e., for both paramagnetic and FM phases and
even at very low $T$. With this in mind, the $E_I$ based
Fermi-Dirac statistics (iFDS) and spin disorder scattering based
resistivity models will be employed in order to derive $\rho$ as a
function of $T$, $E_I$ and $M_{\rho}(T,M_0)$. The consequences of
$\rho(T,E_I,M_{\rho}(T,M_0))$ that arises from the variation of
$T$, $E_I$ and $M_{\rho}(T,M_0)$ are discussed in detail based on
the experimental data reported by Van Esch {\it et
al}.~\cite{esch1}, Mahendiran {\it et al}.~\cite{mahendiran2} and
Park {\it et al}.~\cite{park3}. The Mn$_x$Ge$_{1-x}$ FM
semiconductor (FMS) is also accentuated here due to its promising
properties for device applications~\cite{park3} where its gate
voltage of $\pm$0.5 V is compatible with the present Complementary
Metal-Oxide-Semiconductor (CMOS) and Ge's hole mobility (110.68
m$^2$V$^{-1}$s$^{-1}$) is higher than GaAs (12.65
m$^2$V$^{-1}$s$^{-1}$) and Si (15.81 m$^2$V$^{-1}$s$^{-1}$).
Mn$_x$Ge$_{1-x}$'s resistivity is also semiconductor-like below
$T_C$, which is more suitable than metallic Ga$_{1-x}$Mn$_x$As.
Moreover, Mn$_x$Ge$_{1-x}$ is also the simplest two-element system
that can be utilized to evaluate the performance of the derived
model consists of iFDS and $M_{\alpha}(T)$ (originates from
$\tau_{SD}$). $\alpha$ = K (calculated from the Kasuya's spin
disorder scattering model), $\rho$ (calculated from the
resistivity model), $exp$ (determined experimentally).

\subsection{\textbf{2. Resistivity model}}\lb{s-in}

The total current in semiconducting ferromagnets with
contributions from both paramagnetic and FM phases is $J$ =
$\sum_\nu J_\nu$, $\nu$ = $e^{\downarrow}$, $se^{\uparrow}$,
$h^{\downarrow}$, $sh^{\uparrow}$. For convenience, the spin-up,
$\uparrow$ denotes the direction of the magnetic field or a
particular direction below $T_C$, while the spin-down,
$\downarrow$ represents any other directions. Note that the total
energy (Kinetic + Magnetic), $E_{K+M}^{\uparrow}$ $\neq$
$E_{K+M}^{\downarrow}$ due to energy level splitting below $T_C$.
As such, the total current can be simplified as $J$ =
$J^{\downarrow}_e$ + $J^{\uparrow}_{se}$ = $J_e$ + $J_{se}$ if the
considered system is an $n$-type while $J$ = $J_h$ + $J_{sh}$ if
it is a $p$-type. $J_e$ and $J_h$ are the spin independent charge
current (electrons and holes respectively) in the paramagnetic
phase whereas $J_{se}$ and $J_{sh}$ are the spin-assisted charge
current in the FM phase. Thus the total resistivity ($n$ or
$p$-type) can be written as

\begin {eqnarray}
&\rho^{-1}& = \rho^{-1}_{e,h} + \rho^{-1}_{se,sh} \nn \\&& =
\bigg[\frac{m_{e,h}^*}{(n,p)e^2\tau_e}\bigg]^{-1} +
\bigg[\frac{m_{e,h}^*}{(n,p)e^2\tau_{SD}}\bigg]^{-1}.\label{eq:4}
\end {eqnarray}

$\tau_{SD}$ represents the spin disorder scattering rate. The
carrier density for the electrons and holes ($n,p$) based on iFDS
are given
by~\cite{arulsamy6,arulsamy10,arulsamy11,arulsamy12,arulsamy13}

\begin{eqnarray}
n = 2\left[\frac{k_BT}{2\pi\hbar^2}\right]^{3/2}(m^*_e)^{3/2}
\exp\left[\frac{E_F - E_I}{k_BT}\right]. \label{eq:5}
\end{eqnarray}

\begin{eqnarray}
p = 2\left[\frac{k_BT}{2\pi\hbar^2}\right]^{3/2}(m^*_h)^{3/2}
\exp\left[\frac{-E_F - E_I}{k_BT}\right]. \label{eq:6}
\end{eqnarray}

iFDS is derived in a latter section in which, its applications are
well documented in the
Refs.~\cite{arulsamy6,arulsamy10,arulsamy11,arulsamy12,andrew,andrew2,arulsamy13,andrew3}.
Substituting $1/\tau_e$ = $AT^2$ (due to electron-electron
interaction), Eqs.~(\ref{eq:3}) and~(\ref{eq:5}) or~(\ref{eq:6})
into Eq.~(\ref{eq:4}), then one can arrive at

\begin{eqnarray}
\rho_{e,se}(T) = \frac{AB\exp
\big[(E_I+E_F)/k_BT\big]}{AT^{3/2}[M_{\rho}(T,M_0)]^{-1}+
BT^{-1/2}}. \label{eq:7}
\end{eqnarray}

In which, $A =
[A_{e,h}/2e^2(m^*_{e,h})^{1/2}][2\pi\hbar^2/k_B]^{3/2}$, $B =
2m^*_{e,h}N(\pi E_F)^{1/2}J_{ex}^2/e^2\hbar k_B^{3/2}$ and
$\tau_{SD}^{-1} = [N(2E_F)^{1/2}(m^*_{e,h})^{3/2}/\pi
\hbar^4]J_{ex}^2M_{\rho}(T,M_0)$. $A_{e,h}$ is the $T$ independent
electron-electron scattering rate constant. The $E_I$ here takes
care of the polaronic effect or more precisely, the
electron-phonon interaction. The empirical function of the
normalized magnetization is given by

\begin{eqnarray}
M_{\rho}(T,M_0) = 1-\frac{M_{\rho}(T)}{M_0}\label{eq:8}.
\end{eqnarray}

Equation~(\ref{eq:8}) is an empirical function that directly
quantifies the influence of spin alignments in the FM phase on the
transport properties of charge and spin carriers in accordance
with Eq.~(\ref{eq:7}). In other words, the only way to obtain
$\frac{M_{\rho}(T)}{M_0}$ is through Eq.~(\ref{eq:8}). In fact,
Eq.~(\ref{eq:8}) is used to calculate $M_{TD}(T)/M_0$ and
$M_{K}(T)/M_0$ by writing $S(S+1) -
S^2\big(\frac{M_{TD}(T)}{M_0}\big)^2 -
S\big(\frac{M_{TD}(T)}{M_0}\big)\tanh\big[\frac{3T_CM_{TD}(T)}{2TS(S+1)M_0}\big]$
= $M_{\rho}(T,M_0)$ and $S(S+1) -
S^2\big(\frac{M_{K}(T)}{M_0}\big)^2 -
S\big(\frac{M_{K}(T)}{M_0}\big)$ = $M_{\rho}(T,M_0)$ respectively.
Consequently, one can actually compare and analyze the
$M_{\alpha}(T)/M_0$ ($\alpha$ = TD, K, $\rho$) calculated from
Tinbergen-Dekker (TD), Kasuya (K) and Eq.~(\ref{eq:7}) with the
experimentally measured $M_{exp}(T)/M_0$. However, one has to
switch to Eq.~(\ref{eq:9}) given below for the hole-doped strongly
correlated paramagnetic semiconductors, which is again based on
iFDS~\cite{arulsamy6,andrew},

\begin{eqnarray}
\rho_h =
\frac{A_h(m^*_h)^{\frac{-1}{2}}}{2e^2}\left[\frac{2\pi\hbar^2}{k_B}\right]^{3/2}T^{1/2}
\exp\left[\frac{E_I + E_F}{k_BT}\right]. \label{eq:9}
\end{eqnarray}

$A_h$ is the $T$ independent electron-electron scattering rate
constant. Equation~(\ref{eq:9}) will be used to justify the
importance of the term $J_{se}$ even if the resistivity is
semiconductor-like in the FM phase. Note that, $m^*$ = $m^*_e$
$\approx$ $m^*_h$ $\approx$ $(m^*_em^*_h)^{1/2}$ is used for
convenience. If however, $m^*_e$ $\neq$ $m^*_h$, then one just has
to use the relation, $m^*$ = $m^*_em^*_h/(m^*_e + m^*_h)$. Even in
the usual consideration for the total conductivity, $\sigma$ =
$\sigma_{electron}$ + $\sigma_{hole}$, some algebraic
rearrangements can lead one to the relation, $\rho(T)$ $\propto$
$\exp(E_I/k_BT)/[\exp(E_F/k_BT) + \exp(-E_F/k_BT)]$, exposing the
consistent effect of $E_I$ on transport properties.

\subsection{\textbf{3. Discussion}}\lb{s-in}

\subsection{\textit{3.1. Temperature-dependent resistivity curves}}\lb{s-in}

Resistivity versus temperature measurement ($\rho(T)$) is the most
simplest and effective method to study the transport properties.
In free-electron metals, the $\rho(T)$ curves are often exploited
in order to deduce the $T$-dependence of the scattering rates
namely, $\tau_{e-e}$ and $\tau_{e-ph}$. Such behavior are well
described by the Bloch-Gr\"{u}neisen (BG) formula~\cite{tu1},
given by

\begin{eqnarray}
\rho_{BG} &=& \lambda_{tr} \frac{128 \pi m^*
k_BT^5}{ne^2\Theta_D^4}\int\limits_0^{\Theta_D/2T} \frac
{x^5}{\sinh^2x} dx. \label{eq:10}
\end{eqnarray}

$\lambda_{tr}$ $=$ electron-phonon coupling constant, $n$ $=$ free
electrons concentration. The approximation of $\tau_{e-e}(T)$ and
$\tau_{e-ph}(T)$ using Eq.~(\ref{eq:10}) is valid basically
because there are no other parameters that vary with $T$, apart
from the said scattering rates. In fact, by utilizing the BG
formula, one can reliably estimate that $\tau_{e-e}(T)$ $\propto$
$T^{-2}$ while $\tau_{e-ph}(T)$ $\propto$ $T^{-3\to-5}$ for any
experimentally viable $\Theta_D$.

On the other hand, the metallic phenomenon observed in the
ferromagnetic-metallic (FMM) phase below $T_C$ in ferromagnets
(FM) cannot be characterized as Fermi gas. Therefore, it is rather
incorrect to extract $\tau_{e-e}(T)$, $\tau_{e-ph}(T)$ and
$\tau_{magnons}(T)$ from the $\rho(T<T_C)$ curves in FM.
Experimental evidences based on the photoemission, X-ray
emission/absorption and extended X-ray emission fine structure
spectroscopy have exposed the polaronic effect even at $T$ $<$
$T_C$ or in the FMM phase~\cite{man2}. Consequently, the charge
density ($n$) in FMM phase is not $T$ independent as one would
anticipate for the free-electron metals (Fermi-gas). In addition,
spin related mechanisms, like magnons and spin disorder scattering
can be correctly represented with the normalized magnetization
function, $M(T,M_0)$. It is quite common to employ Matthiessen's
rule ($\tau^{-1} = \sum_i \tau^{-1}_i$) as opposed to the total
current rule ($\tau = \sum_i \tau_i$) and write the resistivity
below $T_C$ in the form of

\begin{eqnarray}
\rho(T) &=& \rho_0 + \sum_i A_iT^{\alpha_i}. \label{eq:11}
\end{eqnarray}

The $i$ here indicates the types of $T$-dependent scattering rate
that contribute to the resistivity and $A$ is a $T$ independent
constant. $\rho_0$ is the $T$-independent scattering rate that
originates from the impurities as $T$ $\to$ 0 K. The critical
issue here is not about the Matthiessen's rule, but on the
validity of Eq.~(\ref{eq:11}) in non free-electronic phase.
Importantly, the $T$-dependent structure of Eq.~(\ref{eq:10}) is
equivalent to Eq.~(\ref{eq:11}) that actually have enabled one to
reliably calculate $\tau^{-1}_{e-e}(T)$ and $\tau^{-1}_{e-ph}(T)$
as $A_{e-e}T^{2}$ and $A_{e-ph}T^{3\to5}$ respectively.
Equation~(\ref{eq:11}) is extremely popular and it is applied
indiscriminately to determine the $T$-dependence of a wide variety
of scattering rates in FMM phase, while the correctness of such
determination is still unclear and varies from one researcher to
another~\cite{mahendiran2,banerjee3,banerjee4}. It is important to
realize that only a free-electronic FMM phase at $T$ $<$ $T_C$
will justify the analysis based on Eq.~(\ref{eq:11}). The
influence of polaronic effect and magnetization function (the
variation of $M(T)/M_0$ with $T$) reinforces the $T$-dependence of
charge density, which point towards the inapplicability of
Eq.~(\ref{eq:11}) in FMM phase.

\subsection{\textit{3.2. Ga$_{1-x}$Mn$_x$As}}\lb{s-in}

The resistivity measurements~\cite{esch1} and its fittings based
on Eqs.~(\ref{eq:7}) and~(\ref{eq:9}) are shown in
Fig.~\ref{fig:1} a) and b) respectively for Ga$_{1-x}$Mn$_x$As.
One needs two fitting parameters ($A$ and $E_I$) for $\rho(T>T_C)$
and another two ($B$ and $M_{\rho}(T,M_0)$) for $\rho(T<T_C)$. All
the fitting parameters are listed in Table~\ref{Table:I}. Note
that $S$ = 1 and 5/2 are employed for the fittings of $M_K(T)/M_0$
while $T_C$ and $T_{crossover}$ = $T_{cr}$ were determined from
the experimental resistivity curves. The deviation of $M_K(T)/M_0$
from the $M_{exp}/M_0$ increases with $S$ from 1 $\to$ 5/2. The
$\rho(T)$ is found to increase with $x$ from 0.060 to 0.070 due to
the mechanism proposed by Van Esch {\it et al.}~\cite{esch1,esch2}
and Ando {\it et al.}~\cite{ando12}. They proposed that neutral
Mn$^{3+}$ acceptors that contribute to magnetic properties could
be compensated by As, where for a higher concentration of Mn,
instead of replacing Ga it will form a six-fold coordinated
centers with As (Mn$^{6As}$)~\cite{esch1,esch2,ando12}. These
centers will eventually reduce the magnitude of ferromagnetism
(FM) in DMS due to the loss of spin-spin interaction between
Mn(3$d^{5})$ and $h$. iFDS based resistivity models also predicts
that if one assumes Mn$^{2+}$ ($E_I$ = 1113 kJmol$^{-1}$) or
Mn$^{3+}$ ($E_I$ = 1825 kJmol$^{-1}$) substitutes Ga$^{3+}$ ($E_I$
= 1840 kJmol$^{-1}$), then $\rho(T)$ should further decrease with
$x$, which is not the case here. Thus, iFDS also suggests that
Mn$^{2+}$ or Mn$^{3+}$ do not substitute Ga$^{3+}$. Interestingly,
the $T_{cr}$s observed in Ga$_{0.940}$Mn$_{0.060}$As (annealed:
370$^o$C) and Ga$_{0.930}$Mn$_{0.070}$As (as grown) are 10 K and
12 K, which are identical with the calculated values of 8 K and 12
K respectively. Note here that $E_I$ + $E_F$ = $T_{cr}$. The
calculated carrier density using $E_I$ + $E_F$ (8, 12 K), $m^*_h$
= rest mass and Eq.~(\ref{eq:6}) is 2.4 $\times$ 10$^{19}$
cm$^{-3}$. Below $T_C$, spin alignments enhance the contribution
from $J_{se}$ and reduces the exponential increase of $\rho(T)$.
This reduction in $\rho(T)$ is as a result of dominating $J_{se}$
and the small magnitude of $E_I$ + $E_F$ (8 K,12 K), consequently
its effect only comes at low $T$ as clearly shown in
Fig.~\ref{fig:1} a). The Ga$_{0.930}$Mn$_{0.070}$As samples after
annealing at 370 $^o$C and 390 $^o$C do not indicate any
FM~\cite{esch1} (Fig.~\ref{fig:1} b)). Thus the fittings are
carried out with Eq.~(\ref{eq:9}) that only requires two fitting
parameters namely, $A$ and $E_I$ + $E_F$ since $J_{se}$ = 0 (there
is no observable $T_C$) and/or $dM_{\alpha}(T)/M_0dT$ = 0
($M_{\rho}(T,M_0)$ = constant). The exponential increase in
Fig.~\ref{fig:1} b) for $\rho(T)$ is due to $E_I$ + $E_F$ from
Eq.~(\ref{eq:9}) with zilch $J_{se}$ contribution.

Figure~\ref{fig:1} c) and d) indicate the normalized
magnetization, $M_{\alpha}(T)/M_0$. Note that
$M_{\rho,TD,K}(T)/M_0$ is a fitting parameter that has been varied
accordingly to fit $\rho(T<T_C)$. $M_{\rho}(T,M_0)$ is used to
calculate $M_{\rho,TD,K}(T)/M_0$ with $S$ = 1.
$M_{\rho,TD,K}(T)/M_0$ is also compared with the experimentally
determined~\cite{esch1} $M_{exp}(T)/M_0$ as depicted in
Fig.~\ref{fig:1} d). One can easily notice the inequality,
$M_{TD}(T)/M_0$ $>$ $M_{K}(T)/M_0$ $>$ $M_{\rho}(T)/M_0$ $>$
$M_{exp}(T)/M_0$ from Fig.~\ref{fig:1} c) and d). As such,
$M_{\rho}(T)/M_0$ from Eq.~(\ref{eq:7}) is the best fit for the
experimentally measured $M_{exp}(T)/M_0$. However,
$M_{\rho}(T)/M_0$ is still larger than $M_{exp}(T)/M_0$, because
resistivity measures only the path with relatively lowest $E_I$
and with easily aligned spins that complies with the principle of
least action. On the contrary, the magnetization measurement
quantifies the average of all the spins' alignments.

\subsection{\textit{3.3. La$_{1-x}$Ca$_x$MnO$_3$}}\lb{s-in}

Mahendiran {\it et al}.~\cite{mahendiran2} discussed $\rho(T<T_C)$
with respect to Eq.~(\ref{eq:1}) and obtained the activation
energy, $E_a$ = 0.16 eV for $x$ = 0.1 and 0.2 of
La$_{1-x}$Ca$_x$MnO$_3$ samples at 0 T. Using Eq.~(\ref{eq:7})
however, $E_I$ + $E_F$ for the former and latter samples are
calculated to be 0.12 and 0.11 eV respectively. The calculated
carrier density using $E_I$ + $E_F$ (0.12, 0.11 eV), $m^*_h$ =
rest mass and Eq.~(\ref{eq:6}) is approximately 10$^{17}$
cm$^{-3}$. In the presence of {\bf H} = 6 T, $E_I$ + $E_F$ is
computed as 0.0776 eV for $x$ = 0.2 that subsequently leads to $p$
= 10$^{18}$ cm$^{-3}$. It is proposed that the activated behavior
for $\rho(T>T_C)$ is due to electron-phonon interaction or rather
due to the polaronic effect ($E_I$)~\cite{arulsamy6}. The fittings
are shown in Fig.~\ref{fig:2} a) and b) while its fitting
parameters are listed in Table~\ref{Table:I}.
Theoretically~\cite{arulsamy6}, Ca$^{2+}$ ($E_I$ = 868
kJmol$^{-1}$) $<$ La$^{3+}$ ($E_I$ = 1152 kJmol$^{-1}$), therefore
$\rho(T)$ is expected to decrease with Ca$^{2+}$ doping
significantly. On the contrary, only a small difference of $E_I$ +
$E_F$ between $x$ = 0.1 (0.12 eV) and 0.2 (0.11 eV) is observed
due to Mn$^{4+}$'s compensation effect. The quantity, Mn$^{4+}$
increased 6\% from $x$ = 0.1 (19\%) to 0.2
(25\%)~\cite{mahendiran2}. Ideally, the difference of $E_I$
between Ca$^{2+}$ and La$^{3+}$ is 1152 $-$ 868 = 284
kJmol$^{-1}$. As a result of compensation with 6\% Mn$^{3+ \to
4+}$ ($E_I$ = 4940 kJmol$^{-1}$), the actual difference is only
284 $-$ \big[0.81(1825) + 0.19(4940) $-$ 0.75(1825) $-$
0.25(4940)\big] = 97 kJmol$^{-1}$. This calculation simply exposes
the compensation effect.

At 6 T, La$_{0.8}$Ca$_{0.2}$MnO$_3$ indicates a much lower
resistivity (Fig.~\ref{fig:2} b)). The result that larger {\bf H}
giving rise to overall conductivity is due to relatively large
amount of aligned spins that eventually gives rise to $J_{se}$.
Hence, $E_I$ + $E_F$ at 6 T (78 meV) is less than $E_I$ + $E_F$ at
0 T (112 meV). Figure~\ref{fig:2} c) and d) depict the calculated
$M_{\alpha}(T)/M_0$ with $S$ = 1 and $M_{exp}(T)/M_0$ for $x$ =
0.2 respectively. The calculated $M_{TD}(T)/M_0$ is dropped for
La$_{1-x}$Ca$_x$MnO$_3$ since $M_K(T)/M_0$ seems to be a better
approximation than $M_{TD}(T)/M_0$ as indicated in
Fig.~\ref{fig:1} c) and d). The discrepancy between
$M_{\rho}(T)/M_0$ and $M_{exp}(T)/M_0$ still exists even though
Eq.~(\ref{eq:7}) reproduces $\rho(T)$ at all $T$ range accurately.
Again, this incompatibility is due to the principle of least
action as stated earlier. In addition, the manganites' charge
transport mechanism below $T_C$ is also in accordance with
Eq.~(\ref{eq:7}) because the term, $M_{\rho}(T,M_0)$ handles the
exchange interactions' complexities separately for DMS and
manganites. For example, one can clearly notice the different type
of discrepancies between DMS and manganites by comparing the
empirical function, $M_{\alpha}(T)/M_0$ ($\alpha$ = $\rho$, exp)
between Fig.~\ref{fig:1} d) and Fig.~\ref{fig:2} d). Hence,
Eq.~(\ref{eq:7}) is suitable for both types of ferromagnets, be it
diluted or concentrated.

\subsection{\textit{3.4. Mn$_x$Ge$_{1-x}$}}\lb{s-in}

The Mn$_x$Ge$_{1-x}$ FMS with homogeneous Mn concentration has
been grown using low-$T$ MBE technique~\cite{park3}. The
Mn$_x$Ge$_{1-x}$ was found to be a $p$-type with carrier density
in the order of $10^{19}-10^{20}$ cm$^{-3}$ for 0.006 $\leq$ $x$
$\leq$ 0.035 as measured by Park {\it et al}.~\cite{park3}. Both
the resistivity measurements~\cite{park3} and its fittings based
on Eq.~(\ref{eq:7}) are shown in Fig.~\ref{fig3} a). Here,
$E_I+E_F$, $A$ and $B$ have been floated while $M_{\rho}(T,M_0)$
is constrained to reduce with $T$ in order to fit the experimental
$\rho(T,x=0.02)$. The absence of the Curie-Weiss law in the
$\rho(T,x=0.02)$ curve is due to insufficient number of aligned
spins that eventually leads to a relatively small $J_{se}$, which
in turn, is not able to produce the metallic conduction below
$T_C$. This scenario is also in accordance with the measured
$M_{exp}(T)$ where, only 1.4-1.9 $\mu_B$/Mn atom contributes to
ferromagnetism as compared with the ideal value of 3.0 $\mu_B$/Mn
atom. In other words, only 45-60$\%$ of Mn ions are magnetically
active~\cite{park3}. It is found that $E_I+E_F$ = 15 K from the
$\rho(T)$ fitting for Mn$_{0.02}$Ge$_{0.98}$. Subsequently, one
will be able to calculate the hole concentration as 2.38 $\times$
10$^{19}$ cm$^{-3}$ using Eq.~(\ref{eq:6}) and $m_h^*$ = rest
mass, which is remarkably in the vicinity of the experimental
value~\cite{park3}, 10$^{19}$ - 10$^{20}$ cm$^{-3}$.
Interestingly, the semiconductor-like behavior of $\rho(T,x=0.02)$
below $T_C$ is {\it not} exponentially driven as the value of
$E_I+E_F$ is very small (15 K) to be able to contribute
significantly above 15 K. Rather, it is the $T$-dependence of
Eq.~(\ref{eq:7}) determines $\rho(T,x=0.02)$ below $T_C$. To see
this clearly, one can actually approximate the experimental
$\rho(T,x=0.02)$ with a mathematical function given by $\rho$ =
$-$21.711 $\times$ $\ln T$ + 148.47 (not shown). In this
computation, the $\ln T$ behavior is the approximate version for
the $T$-dependence of Eq.~(\ref{eq:7}). Another obvious proof is
the inability of Eq.~(\ref{eq:9}) to represent the experimental
$\rho(T,x=0.02)$. The plot using Eq.~(\ref{eq:9}) is also shown in
Fig.~\ref{fig3} a) with $A_h$ = 1.8 and $E_I+E_F$ = 80 K that
eventually give $p$ = 1.92 $\times$ 10$^{19}$ cm$^{-3}$. However,
in the absence of $J_{se}$ term, Eq.~(\ref{eq:9}) is inadequate to
capture the $T$-dependence of $\rho(T,x=0.02)$ in the FM phase.

The pronounced effect of $M_{\rho}(T,M_0)$ can be noticed by
comparing the calculated plots between Eq.~(\ref{eq:7}) and
Eq.~(\ref{eq:7}) with additional constraint, $dM_{\rho}(T)/dT$ = 0
as indicated in Fig.~\ref{fig3} a). Recall that $M_{\rho}(T,M_0)$
is varied with $T$ to fit the experimental $\rho(T,x=0.02)$ in
compliance with Eq.~(\ref{eq:7}). Furthermore, $\rho(T)$ is
found~\cite{park3} to decrease with $x$ from 0.016 to 0.02 while
$\rho(T,x=0.02)$ remains identical with $\rho(T,x=0.033)$. This
type of transition can be readily evaluated with Eq.~(\ref{eq:7}).
Firstly, notice the large increase in room temperature $p$ from
10$^{14}$ cm$^{-3}$ (upper limit) for pure Ge to 10$^{19}$
cm$^{-3}$ (lower limit) for a mere 2$\%$ Mn substituted
Mn$_{0.02}$Ge$_{098}$, which gives rise to a rapid decrease of
$\rho(T,x)$. The average $E_I$s for Mn$^{2+}$, Mn$^{3+}$ and
Ge$^{4+}$ are computed as 1113, 1825 and 2503 kJmol$^{-1}$
respectively. According to iFDS, Mn substitution into Ge sites
will reduce the magnitude of $\rho(T)$ since $E_I$ (Ge$^{4+}$) $>$
$E_I$ (Mn$^{3+}$) $>$ $E_I$ (Mn$^{2+}$), regardless of $dM(T)/dT$
= 0 or $dM(T)/dT$ $\neq$ 0. Such behavior has been observed
experimentally~\cite{park3} where, $\rho(T,x=0.009)$ $>$
$\rho(T,x=0.016)$ $>$ $\rho(T,x=0.02)$. This scenario indicates
that the holes from Mn$^{2+,3+}$ is kinetically favorable than the
intrinsic holes from Ge$^{4+}$. It is also found experimentally
that $\rho_{rt} (x=0.009)$ $<$ $\rho_{rt} (x=0.016)$ as a result
of the variation of $T$ independent scattering rate constants ($A$
and $B$). Surprisingly however, $\rho(T,x=0.02)$ $\simeq$
$\rho(T,x=0.033)$, which suggests that $A$, $B$ and $T$-dependence
of $M_{exp}(T)/M_0$ are identical. Using iFDS, one should get
$\rho(T,x=0.02)$ $>$ $\rho(T,x=0.033)$ and $[A,B]_{x=0.02}$ $>$
$[A,B]_{x=0.033}$. Meaning, the additional Mn substitution (0.033
$-$ 0.02 = 0.013) may not have substituted Ge, instead it could
have formed a well segregated impurity phase that eventually
contributes to the higher magnitudes of $A$ and $B$
($[A,B]_{x=0.02}$ $\approx$ $[A,B]_{x=0.033}$), and consequently
does not interfere with the $M_{exp}(T)/M_0$. Notice that the
formation of impurity phase is quite common in any system,
including Ga$_{1-x}$Mn$_x$As DMS with strictly limited Mn
solubility. On the other hand, the normalized magnetization,
$M_{K,\rho,exp}(T)/M_0$ for Mn$_{0.02}$Ge$_{0.98}$ have been
plotted in Fig.~\ref{fig3} b). One can notice the relation,
$M_{K}(T)/M_0$ $>$ $M_{\rho}(T)/M_0$ $>$ $M_{exp}(T)/M_0$ from
Fig.~\ref{fig3} b). Again, $M_{\rho}(T)/M_0$ $>$ $M_{exp}(T)/M_0$
is due to the ability of both $J_e$ and $J_{se}$ to follow the
easiest path. Additionally, the $T$-dependence of $M_{exp}(T)$ is
similar to Ga$_{1-x}$Mn$_x$As rather than the well established
manganite ferromagnets, which reveals the possibility of multiple
exchange interactions~\cite{park3,esch1,esch2}. All the values of
$E_I$ discussed above were averaged in accordance with $E_I
[X^{z+}] = \sum_{i=1}^z\frac{E_{Ii}}{z}$. Prior to averaging, the
1$^{st}$, 2$^{nd}$, 3$^{rd}$ and 4$^{th}$ ionization energies for
all the elements mentioned above were taken from Ref.~\cite{web}.

\subsection{\textbf{4. iFDS}}

Here, the Lagrange multipliers, degeneracy and the total energy
requirement associated with $E_I$ in iFDS is discussed in detail.
Both FDS and iFDS are for the half-integral spin particles such as
electrons and holes. Its total wave function, $\Psi$ has to be
antisymmetric in order to satisfy quantum-mechanical symmetry
requirement. Under such condition, interchange of any 2 particles
($A$ and $B$) of different states, $\psi_i$ and $\psi_j$ ($j$
$\ne$ $i$) will result in change of sign, hence the wave function
for Fermions is in the form of

\begin {eqnarray}
\Psi_{i,j}(C_A,C_B) = \psi_i(C_A)\psi_j(C_B) -
\psi_i(C_B)\psi_j(C_A). \label{eq:111}
\end {eqnarray}

The negative sign in Eq.~(\ref{eq:111}) that fulfils antisymmetric
requirement is actually due to one of the eigenvalue of exchange
operator~\cite{griffiths6}, {\bf P} = $-$1. The other eigenvalue,
{\bf P} = $+$1 is for Bosons. $C_A$ and $C_B$ denote all the
necessary cartesian coordinates of the particles $A$ and $B$
respectively. Equation~(\ref{eq:111}) is nothing but Pauli's
exclusion principle. The one-particle energies $E_1$, $E_2$, ...,
$E_m$ for the corresponding one-particle quantum states $q_1$,
$q_2$, ..., $q_m$ can be rewritten as ($E_{is}$ $\pm$ $E_I)_1$,
($E_{is}$ $\pm$ $E_I)_2$, ..., ($E_{is}$ $\pm$ $E_I)_m$. Note here
that $E_{is}$ = $E_{initial~state}$. It is also important to
realize that $E_{is}$ + $E_I$ = $E_{electrons}$ and $E_{is}$ $-$
$E_I$ = $E_{holes}$. Subsequently, the latter ($E_{is}$ $\pm$
$E_I)_i$ version where $i$ = 1, 2, ..., $m$ with $E_I$ as an
additional inclusion will be used to derive iFDS and its Lagrange
multipliers. This $\pm E_I$ is inserted carefully to justify that
an electron to occupy a higher state $N$ from initial state $M$ is
more probable than from initial state $L$ if condition $E_I(M)$
$<$ $E_I(L)$ at certain $T$ is satisfied. As for a hole to occupy
a lower state $M$ from initial state $N$ is more probable than to
occupy state $L$ if the same condition above is satisfied.
$E_{is}$ is the energy of a particle in a given system at a
certain initial state and ranges from $+\infty$ to 0 for electrons
and 0 to $-\infty$ for holes. In contrast, standard FDS only
requires $E_i$ ($i$ = 1, 2, ..., $m$) as the energy of a particle
at a certain state.

Denoting $n$ as the total number of particles with $n_1$ particles
with energy ($E_{is}$ $\pm$ $E_I)_1$, $n_2$ particles with energy
($E_{is}$ $\pm$ $E_I)_2$ and so on implies that $n$ = $n_1$ +
$n_2$ + ... + $n_m$. As a consequence, the number of ways for
$q_1$ quantum states to be arranged among $n_1$ particles is given
as

\begin {eqnarray}
P(n_1,q_1) = \frac{q_1!}{n_1!(q_1 - n_1)!}.\label{eq:222}
\end {eqnarray}

Now it is easy to enumerate the total number of ways for $q$
quantum states ($q$ = $q_1$ + $q_2$ + ... + $q_m$) to be arranged
among $n$ particles, which is

\begin {eqnarray}
P(n,q) = \prod\limits_{i=1}^{\infty} \frac{q_i!}{n_i!(q_i - n_i)!}
.\label{eq:333}
\end {eqnarray}

The most probable configuration at certain $T$ can be obtained by
maximizing $P(n,q)$ subject to the restrictive conditions

\begin {eqnarray}
&&\sum_i^{\infty} n_i = n, \sum_i^{\infty} dn_i = 0.
\label{eq:444}
\end {eqnarray}

\begin {eqnarray}
&&\sum_i^{\infty} (E_{is}\pm E_I)_i n_i = E, \sum_i^{\infty}
(E_{is}\pm E_I)_i dn_i = 0.\label{eq:555}
\end {eqnarray}

The method of Lagrange multipliers~\cite{griffiths6} can be
employed to maximize Eq.~(\ref{eq:333}). Hence, a new function,
$F(x_1, x_2, ...\mu, \lambda,...)$ = $f + \mu f_1 + \lambda f_2$
+... is introduced and all its derivatives are set to zero

\begin {eqnarray}
\frac{\partial F}{\partial x_n} = 0;~~~ \frac{\partial F}{\partial
\mu} = 0;~~~ \frac{\partial F}{\partial \lambda} =
0.\label{eq:666}
\end {eqnarray}

As such, one can let the new function in the form of

\begin {eqnarray}
F = \ln P + \mu \sum_i^{\infty} dn_i + \lambda \sum_i^{\infty}
(E_{is}\pm E_I)_i dn_i.\label{eq:777}
\end {eqnarray}

After applying Stirling's approximation, $\partial F$/$\partial
n_i$ can be written as

\begin {eqnarray}
\frac {\partial F}{\partial n_i}&& = \ln (q_i - n_i) - \ln n_i +
\mu + \lambda (E_{is}\pm E_I)_i \nn \\&& = 0.\label{eq:888}
\end {eqnarray}

Thus, the Fermi-Dirac statistics based on ionization energy is
simply given by

\begin {eqnarray}
\frac {n_i}{q_i} = \frac{1}{\exp [\mu + \lambda (E_{is}\pm E_I)_i]
+ 1}.\label{eq:999}
\end {eqnarray}


Importantly, the total energy, $E$ in iFDS can be obtained from
Eq.~(\ref{eq:555}), which is

\begin {eqnarray}
&E& = \sum_i^{\infty} (E_{is}\pm E_I)_i n_i \nn \\&& =
\sum_i^{\infty} \frac{\hbar^2}{2m}\big[{\bf k}^2_{is}\pm {\bf
k}^2_I\big]_i n_i \nn
\\&& = \frac{\hbar^2}{2m}\big[{\bf k}^2_{is}\pm {\bf
k}^2_I\big] = \frac{\hbar^2}{2m}{\bf k}^2.\label{eq:100}
\end {eqnarray}

$\textbf{k}_I$ = $\textbf{k}_{ionized~state}$, and the $\pm$ sign
is solely to indicate that the energy corresponds to electrons is
0 $\to$ $+\infty$ while 0 $\to$ $-\infty$ is for the holes, which
satisfy the particle-hole symmetry. Consequently,
Eq.~(\ref{eq:100}) also implies that iFDS does not violate the
degeneracy requirements. By utilizing Eq.~(\ref{eq:999}) and
taking $\exp[\mu + \lambda(E \pm E_I)]$ $\gg$ 1, one can arrive at
the probability function for electrons in an explicit form as

\begin{eqnarray}
f_e({\bf k}_{is}) = \exp \left[-\mu-\lambda\left(\frac{\hbar^2{\bf
k}_{is}^2}{2m}+E_I\right) \right], \label{eq:110}
\end{eqnarray}

Similarly, the probability function for the holes is given by

\begin{eqnarray}
f_h({\bf k}_{is}) = \exp\left[\mu + \lambda\left(\frac{\hbar^2{\bf
k}_{is}^2}{2m}-E_I\right) \right]. \label{eq:120}
\end{eqnarray}

The parameters $\mu$ and $\lambda$ are the Lagrange multipliers.
$\hbar$ $=$ $h/2\pi$, $h$ $=$ Planck constant and $m$ is the
charge carriers' mass. Note that $E$ has been substituted with
$\hbar^2{\bf k}^2/2m$. In the standard FDS, Eqs.~(\ref{eq:110})
and~(\ref{eq:120}) are simply given by, $f_e({\bf k})$ $=$
$\exp[-\mu-\lambda(\hbar^2{\bf k}^2/2m)]$ and $f_h({\bf k})$ $=$
$\exp[\mu+\lambda(\hbar^2{\bf k}^2/2m)]$. Equation~(\ref{eq:444})
can be rewritten by employing the 3D density of states' (DOS)
derivative, $dn$ $=$ $V{\bf k}_{is}^2d{\bf k}_{is}/2\pi^2$,
Eqs.~(\ref{eq:110}) and~(\ref{eq:120}), that eventually give

\begin{eqnarray}
&n& = \frac {V}{2\pi^2}e^{-\mu} \int\limits_0^\infty {\bf k}^2
\exp\bigg[-\lambda \frac{\hbar^2{\bf k}^2}{2m}\bigg] d{\bf k} \nn
\\&& = \frac {V}{2\pi^2}e^{-\mu} \int\limits_0^\infty {\bf
k}_{is}^2 \exp\bigg[-\lambda \frac{\hbar^2{\bf k}_{is}^2}{2m}
-\lambda \frac{\hbar^2{\bf k}_I^2}{2m} \bigg] d{\bf k}_{is} \nn
\\&& = \frac {V}{2\pi^2}e^{-\mu-\lambda E_I} \int\limits_0^\infty
{\bf k}_{is}^2 \exp\bigg[-\lambda \frac{\hbar^2{\bf
k}_{is}^2}{2m}\bigg] d{\bf k}_{is}, \label{eq:130}
\end{eqnarray}

\begin{eqnarray}
p & = & \frac {V}{2\pi^2}e^{\mu - \lambda E_I}
\int\limits_{-\infty}^0 {\bf k}_{is}^2 \exp\bigg[\lambda \frac
{\hbar^2{\bf k}_{is}^2}{2m}\bigg] d{\bf k}_{is}.\label{eq:140}
\end{eqnarray}

The respective solutions for Eqs.~(\ref{eq:130})
and~(\ref{eq:140}) are

\begin{eqnarray}
\mu + \lambda E_I & = &
-\ln\left[\frac{n}{V}\left(\frac{2\pi\lambda\hbar^2}{m}\right)^{3/2}
\right],\label{eq:150}
\end{eqnarray}

\begin{eqnarray}
\mu -\lambda E_I & = &
\ln\left[\frac{p}{V}\left(\frac{2\pi\lambda\hbar^2}{m}\right)^{3/2}
\right].\label{eq:160}
\end{eqnarray}

Note that Eqs.~(\ref{eq:150}) and~(\ref{eq:160}) simply imply that
$\mu_e(iFDS)$ $=$ $\mu(T=0)$ + $\lambda E_I$ and $\mu_h(iFDS)$ $=$
$\mu(T=0)$ $-$ $\lambda E_I$. In fact, $\mu(FDS)$ need to be
varied accordingly with doping, on the other hand, iFDS captures
the same variation due to doping with $\lambda E_I$ in which,
$\mu(T=0)$ is fixed to be a constant (independent of $T$ and
doping). Furthermore, using Eq.~(\ref{eq:555}), one can obtain

\begin{eqnarray}
E && = \frac {V\hbar^2}{4m\pi^2} e^{-\mu(FDS)}
\int\limits_0^\infty {\bf k}^4 \exp\bigg[-\lambda
\frac{\hbar^2{\bf k}^2}{2m}\bigg] d{\bf k} \nn
\\&& = \frac {V\hbar^2}{4m\pi^2} e^{-\mu(T=0)} \int\limits_0^\infty {\bf
k}_{is}^4 \exp\bigg[-\lambda \frac{\hbar^2{\bf k}_{is}^2}{2m}
-\lambda \frac{\hbar^2{\bf k}_I^2}{2m} \bigg] d{\bf k}_{is} \nn
\\&& = \frac {V\hbar^2}{4m\pi^2} e^{-\mu(T=0) -\lambda E_I}
\int\limits_0^\infty {\bf k}_{is}^4
\exp\bigg[-\lambda\frac{\hbar^2{\bf k}_{is}^2}{2m}\bigg]d{\bf
k}_{is} \nn
\\&& = \frac{3V}{2\lambda}e^{-\mu(T=0) -\lambda E_I}
\bigg[\frac{m}{2\pi\lambda\hbar^2}\bigg]^{3/2} \label{eq:170}
\\&& = \frac{3V}{2\lambda}e^{-\mu(FDS)}
\bigg[\frac{m}{2\pi\lambda\hbar^2}\bigg]^{3/2}. \label{eq:180}
\end{eqnarray}

Again, Eq.~(\ref{eq:170}) being equal to Eq.~(\ref{eq:180}) enable
one to surmise that the total energy considered in FDS and iFDS is
exactly the same. Quantitative comparison between
Eq.~(\ref{eq:170}) and with the energy of a 3D ideal gas, $E$ $=$
$3nk_BT/2$, after substituting Eq.~(\ref{eq:150}) into
Eq.~(\ref{eq:170}) will enable one to determine $\lambda$. It is
found that $\lambda$ remains the same as 1/$k_BT$. Basically, at
constant temperature ($T$ $>$ 0), FDS predicts the distribution
spectrum if $E$ is varied, relying on external inputs such as band
gap ($E_g$) and/or Fermi level ($E_F(T)$). On the other hand, iFDS
needs only $E_I$ as an external input to predict the variation of
$E$, without relying on $E_g$ and/or $E_F(T)$ at all, and
subsequently its distribution spectrum can be obtained as well.
Notice that $E_F$ comes into iFDS as $E_F^0$ = constant,
independent of $T$ and doping. $E_F^0$ denotes the Fermi level at
0 K. In fact, iFDS and FDS take different approach in term of
energy levels and Fermions' excitations to arrive at the same
distribution spectrum. In simple words, iFDS is new in a sense
that it gives one an alternative route to obtain the Fermions'
distribution spectrum in which, FDS needs $E_F(T)$ and/or $E_g$
while iFDS needs only $E_I$ to arrive at the same distribution
spectrum. Hence, based on the accuracy of these input parameters,
one can choose either FDS or iFDS to be used for one's theoretical
models.

\subsection{\textbf{5. Conclusions}}\lb{s-in}

In conclusion, the transport properties of Ga$_{1-x}$Mn$_x$As,
La$_{1-x}$Ca$_x$MnO$_3$ and Mn$_x$Ge$_{1-x}$ can be characterized
with a model consists of ionization energy based Fermi-Dirac
statistics coupled with spin disorder scattering mechanism. This
model has been able to explain the evolution of resistivity's
curves with respect to temperature and Mn doping. The arguments
for the incompatibility between the calculated and experimentally
determined normalized magnetization is based on the total
current's tendency to obey the principle of least action. The
validity of $E_I+E_F$ and $M_{\rho}(T,M_0)$ have been justified
quantitatively by computing $p$ and $M_{\rho}(T)/M_0$
respectively, which are in excellent agreement with the
experimental results. However, the magnitudes of $A$ and $B$ are
not diagnosed due to unknown reliable values of $A_h$, $J_{ex}$
and $E_F$. To this end, the variation of hole mobilities and
dielectric constant with doping, the influence of multiple
exchange interaction and energy gap above $T_C$ should be
investigated experimentally.

\section*{\textbf{Acknowledgments}}

The author is grateful and beholden to Arulsamy Innasimuthu,
Sebastiammal Innasimuthu, Arokia Das Anthony and Cecily Arokiam of
CMG-A for their extended financial aid. ADA also thanks Bryne
J.-Y. Tan, Jasper L. S. Loverio and Hendry Izaac Elim for their
kind help with figure preparations and references.

\begin{figure}

\caption {Equation~(\ref{eq:7}) has been employed to fit the
experimental $\rho(T)$ plots for Ga$_{1-x}$Mn$_x$As as given in a)
whereas Eq.~(\ref{eq:9}) is used to fit the plots in b). All
fittings are indicated with solid lines. b) is actually for
annealed non-ferromagnetic Ga$_{0.930}$Mn$_{0.070}$As samples. c)
and d) show the $T$ variation of calculated
$M_{\alpha}(T)/M_{4.2}$ ($\alpha$ = K, TD, $\rho$) with $S$ = 1
for $x$ = 0.060 and 0.070 respectively. $M_K(T)/M_{4.2}$ is also
calculated with $S$ = 5/2. The experimental $M_{exp}(T)/M_{4.2}$
plot for $x$ = 0.070 (as grown) is shown in d).} \label{fig:1}
\end{figure}

\begin{figure}

\caption {Experimental plots of $\rho(T)$ for
La$_{1-x}$Ca$_x$MnO$_3$ at $x$ = 0.1, 0.2 and 0.2 (6 T) have been
fitted with Eq.~(\ref{eq:6}) as depicted in a) and b). All
fittings are indicated with solid lines. Whereas c) and d) show
the $T$ variation of calculated $M_{\alpha}(T)/M_{4.2}$ ($\alpha$
= K, $\rho$) with $S$ = 1 for $x$ = 0.1 and 0.2 respectively. The
experimental $M_{exp}(T)/M_{4.2}$ plot for $x$ = 0.2 is given in
d).} \label{fig:2}
\end{figure}

\begin{figure}
\caption {a) Equation~(\ref{eq:7}) has been employed to fit the
experimental $\rho(T)$ plots for Mn$_{0.02}$Ge$_{0.98}$. The plot
with additional constraint, $dM_{\rho}(T)/dT$ = 0 on
Eq.~(\ref{eq:7}) is also given to emphasize the influence of
$M_{\rho}(T)/M_0$ for an accurate fitting. In these two plots, $A$
= 25, $B$ = 1060 and $E_I+E_F$ = 15 K. The $T$-dependence of
$\rho(T)$ in accordance with $J_e$ only, ignoring $J_{se}$ is
calculated with Eq.~(\ref{eq:9}), which lacks the ability to
capture the experimental $\rho(T,x=0.02)$. In this case, $A_h$ =
1.8 and $E_I+E_F$ = 15 K. Both $E_I+E_F$ = 15 K and $E_I+E_F$ = 80
K give $p$ in the order of 10$^{19}$ cm$^{-3}$ using
Eq.~(\ref{eq:6}) and $m^*_h$ = rest mass. b) Shows the $T$
variation of $M_{\alpha}(T)/M_0$ ($\alpha = K, \rho, exp)$ for $x$
= 0.02. Notice the inequality, $M_{K}(T)/M_0$ $>$
$M_{\rho}(T)/M_0$ $>$ $M_{exp}(T)/M_0$ that arises as a result of
the principle of least action. The $T$-dependence of
$M_{\alpha}(T)/M_0$ is close to the Ga$_{1-x}$Mn$_x$As DMS, rather
than the traditional manganites. As such, this behavior is
suspected to be associated with the multiple exchange
interaction.} \label{fig3}
\end{figure}

\begin{table}

\caption {Calculated values of $T$ independent electron-electron
scattering rate constant ($A$), $B$, which is a function of $T$
independent spin disorder scattering rate constant and spin
exchange energy ($J_{ex}$) as well as the ionization energy
($E_I$). All these parameters are for Mn doped Ga$_{1-x}$Mn$_x$As
(as grown and annealed at 370 $^o$C, 390 $^o$C) and Ca doped
La$_{1-x}$Ca$_x$MnO$_3$ (measured at 0 and 6 T) systems. All
Ga$_{1-x}$Mn$_x$As samples were measured at 0 T.} \label{Table:I}
\end{table}

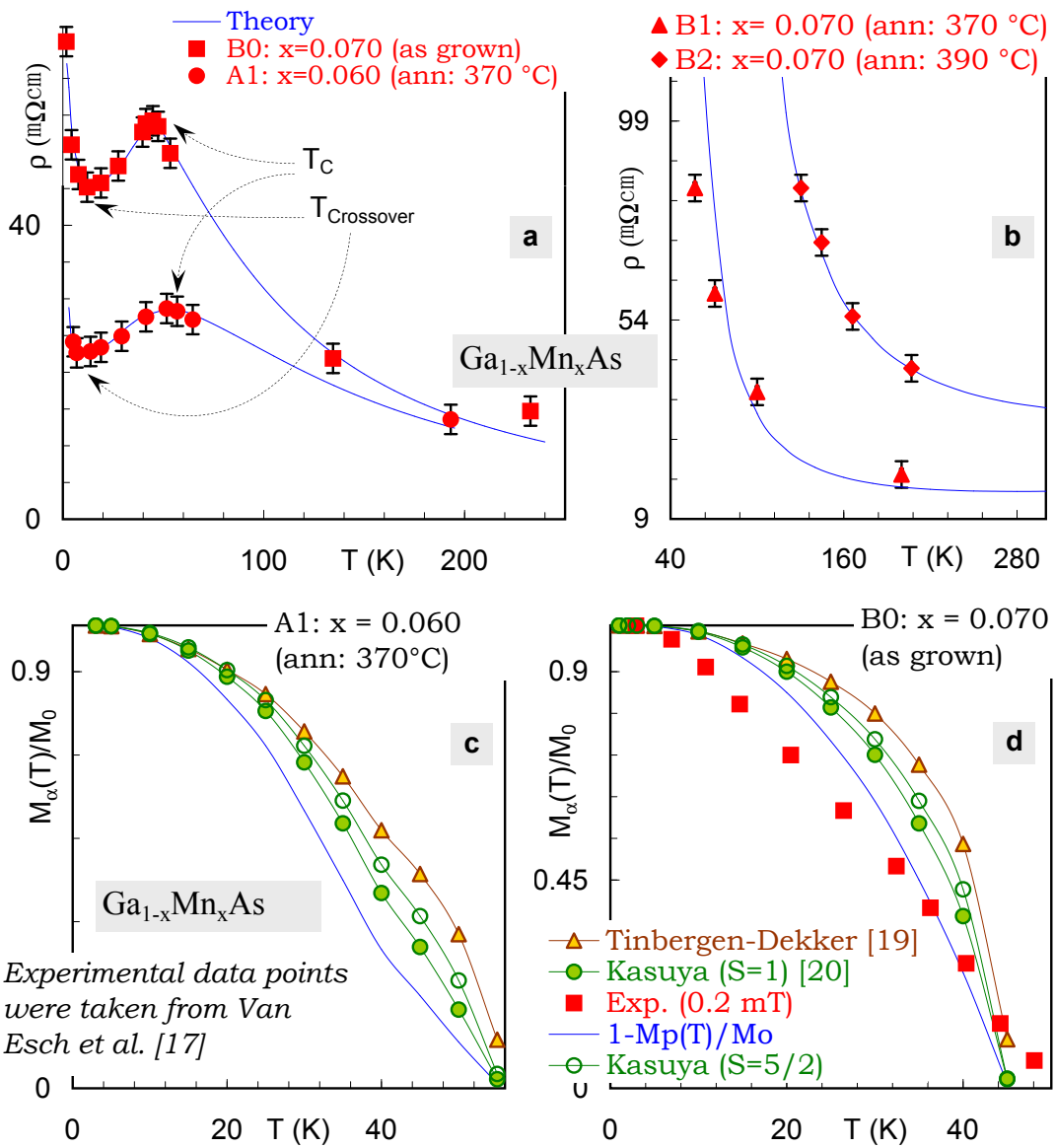
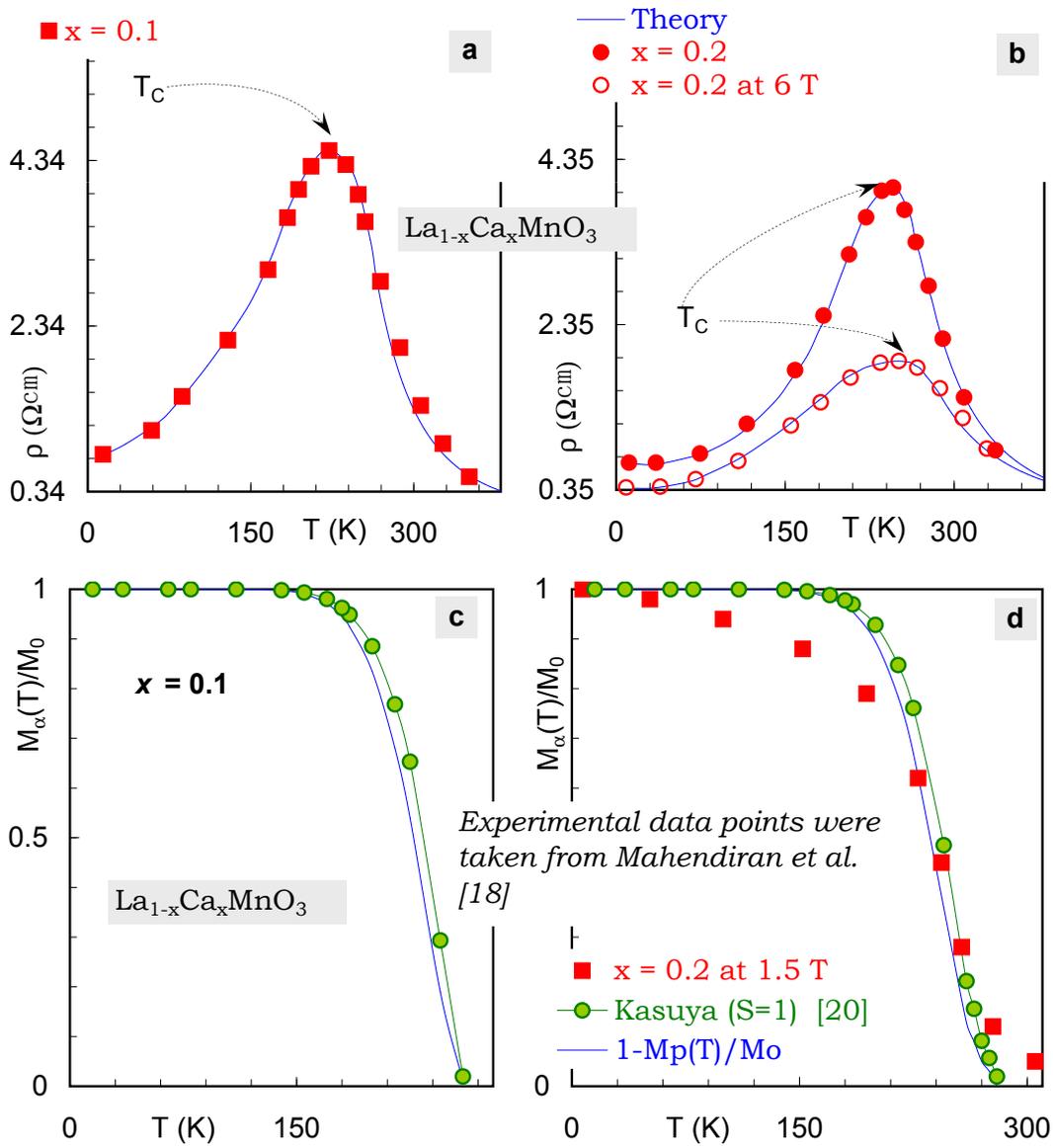
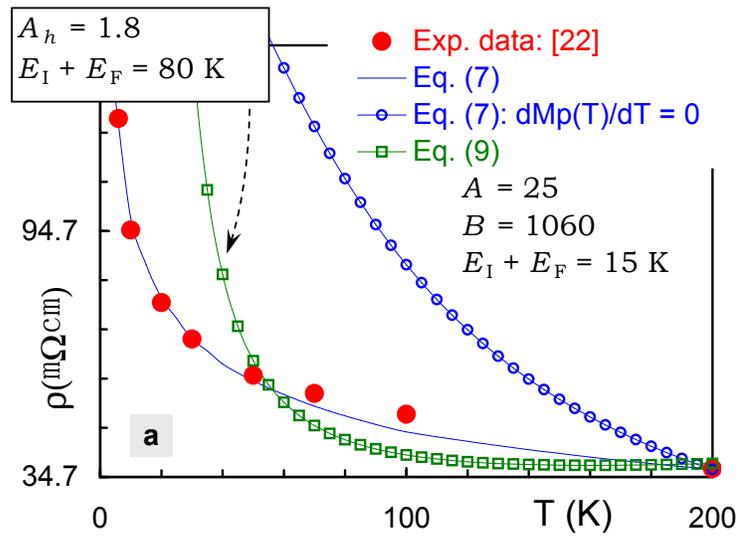
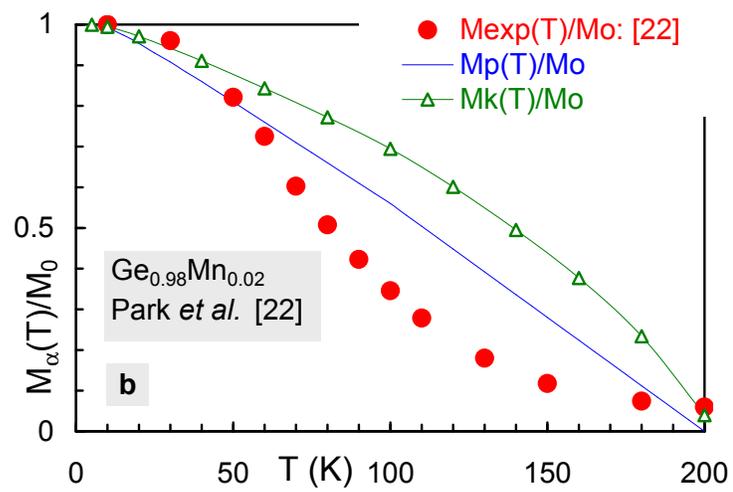
\end{document}